# Magnetoresistance in a soft billiard: giant peak near the percolation threshold

Michel Dyakonov[1] and Rémi Jullien[2]
[1]*Laboratoire de Physique Théorique et Astroparticules*[†], *Université Montpellier 2, France*
[2]*Laboratoire des Colloïdes, Verres et Nanomatériaux*[†], *Université Montpellier 2, France*
[†] *Laboratoire associé au Centre National de la Recherche Scientifique (CNRS, France).*

By numerical simulation, we study the classical magnetoresistance of two-dimensional electrons in the presence of weak short range scattering. A critical magnetic field defines the percolation threshold, above which the electrons the longitudinal resistance vanishes. Unexpectedely, just below this threshold we find a sharp narrow peak, where the resitance may increase 15 times compared to its zero-field value. By considering the complex topology of the effective potential landscape for the center of the cyclotron circle, we show that this phenomenon is related to infinite equipotential lines, which exist only in a narrow magnetic field interval below the percolation threshold.

Along with quantum interference, classical memory effects in transport are beyond the conventional Boltzmann-Drude approach, which does not account for particles recolliding with the same scattering center or repeatedly passing through a space region which is free of such centers. These processes are responsible for the power-law tail in the velocity correlation function and an increase of the resistivity compared to its Boltzmann value [1, 2].

The role of classical memory effects becomes greatly enhanced in the presence of a magnetic field since the circling motion of an electron naturally increases the probability of returns, as it was demonstrated for the first time by Baskin, Magarill, and Entin [3] (see also [4]). They considered scattering by hard disks in two dimensions and showed that, except for the case of low magnetic fields, the Boltzmann-Drude approach does not work, even as a first approximation, because of the existence of "circling" electrons, which never collide with the short range scattering centers. Contrary to the assumption intrinsic to the Boltzmann-Drude approach, an electron which happens to make one collisionless cycle will stay on its cyclotron orbit forever. Interestingly, it took 20 years to recognize that this results in a negative magnetoresistance [5, 6] and that there are not only quantum, but also purely classical reasons for magnetoresistance . [We recall that for the case of elastic scattering of degenerate electrons the Boltzmann approach yealds zero magnetoresistance, i.e. the longidutinal resistivity, $\rho_{xx}$, does not depend on magnetic field. Thus, for a degenerate 2D electron gas, the magnetoresistance is entirely due to memory effects, either classical or quantum].

For the case of scattering by a weak random potential, it was shown [7] that classical localization arises when the magnetic field becomes strong enough to make the shift of the electron orbit during the cyclotron period smaller than the potential correlation lengh. In this regime the center of the cyclotron orbit adiabatically follows the equipotential lines of the averaged potential. This leads to the decrease of the diagonal conductivity and hence to negative magnetoresistance. At lower magnetic fields the intersections of the electron trajectory lead to correlated shifts of the cyclotron orbit and result, on the contrary, in a positive magnetoresistance [8]. It is remarkable that the same mechanism of returns to previously visited regions may lead either to the enhancement of chaotic motion or to localization, depending on whether the regime is adiabatic or not. Returns to the same region due to backscattering (the corridor effect) are responsible for the anomalous magnetoresistance in classicaly weak magnetic fields [9, 10].

An interesting property of short range scattering is the existence of a critical magnetic field corresponding to a percolation threshold [3], above which all electrons get localized and the longitudinal resistance vanishes. The reason is that the trajectory of an electron colliding with an isolated scatterer (of diameter $d$) has the form of a rosette, sweeping a circular area around the scattering center. If the average number of impurities inside this area $\sim NR^2$ is large, where $N$ is the concentration of scatterers and $R$ is the cyclotron radius, eventually the electron will collide with one of them, and thus continue its diffusion in the 2D plane. However, at strong fields, when the parameter $NR^2$ becomes small enough, the rosettes around different scatterers do not overlap anymore, and the colliding electrons become localized. This happens at the percolation threshold for disks of radius $R + d/2$ ($R >> d$), which is known to be $N(R + d/2)^2 = 0.36$ [4, 11].

As a part of a systematic investigation of classical memory effects in magnetotransport [12], in this Letter we present results of numerical simulation, as well as a qualitative interpretation, of magnetoresitance of non-interacting electrons in a dilute system of weak short range scatterers in two dimensions, which one might call a "soft billiard". While it may be possible to approach this situation experimentally by using a weakly doped quantum well with high electron density, we see the main interest of this study in revealing the intricate interplay between the chaotic behavior and localization. The system is characterized by two small parameters: the dimensionless concentration $c = \pi N d^2/4 << 1$ and the typical scattering angle $\alpha = U/E << 1$, where $U$ is the scattering potential, $E$ is the electron energy (the Fermi energy).

Our simulation shows, indeed zero $\rho_{xx}$ above the cor-responding value of magnetic field, however we find that just below the percolation threshold the magnetoresistance exhibits a sharp and narrow peak where the resistance may be 15 times higher than its zero-field value. We qualitatively explain this extraordinary behavior as the consequence of the complex topology of the effective potential landscape for the center of the cyclotron orbit.

In the numerical simulations, we consider a set of disks of diameter $d$ with centers randomly positionned in a square of edge length $L = 4000d$. The total number of disks is chosen to fit the desired concentration. Each disk creates a radial potential

$$u(r) = U\cos^2(\pi r/d) \text{ for } r < d/2;\ u(r) = 0 \text{ otherwise.} \quad (1)$$

This choice is quite arbitrary. However it was checked that qualitatively the results do not depend on the precise form of the potential, so long as it remains smooth near the disk edge, insuring the continuity of the force acting on the electron. As in our previous works[2, 6, 9] we numerically determine the electron trajectories by starting from a random initial position and a randomly oriented initial velocity. When the electron is inside a disk, a standard molecular dynamics method (taking care of the rotation in magnetic field) is used with a time step of $0.02d/v$, where $v$ is the electron velocity outside of the disks. Then the relevant physical quantities are calculated after making an ensemble average over a large number of trajectories. We have checked that, when the electron energy $E$ is smaller than $U$ ($\alpha > 1$), the same results are obtained as for the hard disk (Lorentz) model [2], i.e. a negative magnetoresistance. However in the large energy limit ($\alpha << 1$) the results are completely different. In particular the transport cross-section $\sigma$ becomes very small, of the order $d\alpha^2$, and the Boltzmann mean-free path is given by $\ell = 1/(N\sigma) = 4.857d/(c\alpha^2)$ (the numerical coefficient depends on the exact form of the potential $u(r)$). In this study, we are concerned with magnetic field values such that $\beta = \omega_c\tau \sim \ell/R >> 1$, where $\omega_c$ is the cyclotron frequency and $\tau$ is the transport relaxation time. Under this condition the Hall conductivity is much greater than the diagonal conductivity, $\sigma_{xx} << \sigma_{xy} \simeq \sigma_0/\beta$, where $\sigma_0$ is the Boltzmann conductivity. As a consequence, the longitudinal resistivity, normalized to its zero-field value, is given by:

$$\rho_{xx} = D_{xx}/D_{xx}^B, \quad (2)$$

where $D_{xx}$ is the $xx$ component of the diffusion tensor in magnetic field, $D_{xx}^B = D_0/\beta^2$ is its Boltzmann value at $\beta >> 1$, and $D_0$ is the zero field Boltzmann diffusion coefficient. Here the proportionality between the diffusion and the conductivity tensors is used.

The condition $\beta >> 1$ allows to consider the trajectory of the center of the cyclotron orbit, rather than the electron trajectory itself [7]. So, instead of calculating $D_{xx}$ through the integral of the velocity-velocity correlator as it was done previously [2, 6, 9], we now find this quantity directly from the diffusive trajectory of the cyclotron circle center. At each time step of the molecular dynamics, we determine its coordinates $x_c$ and $y_c$ and find $D_{xx}$ by looking at the large time limit of $(x_c^2 + y_c^2)/t$. To establish this limit correctly, especially for magnetic fields in the vicinity of the percolation threshold, it is necessary to run up to a very large maximum time $t_m$. We use the value $t_m = 500\tau$, which was found to be sufficient to ensure reliable convergence for $D_{xx}$. The advantage of this method is that it eliminates large fluctuations due to cyclotron oscillations of the electron velocity components. Thus one can use a considerably smaller number of trajectories in the ensemble average to obtain a comparable precision for the results. Here we have considered 100 independent disk configurations and 100 independant starting conditions for each configuration. We have checked that this method gives the same results as the previous one.

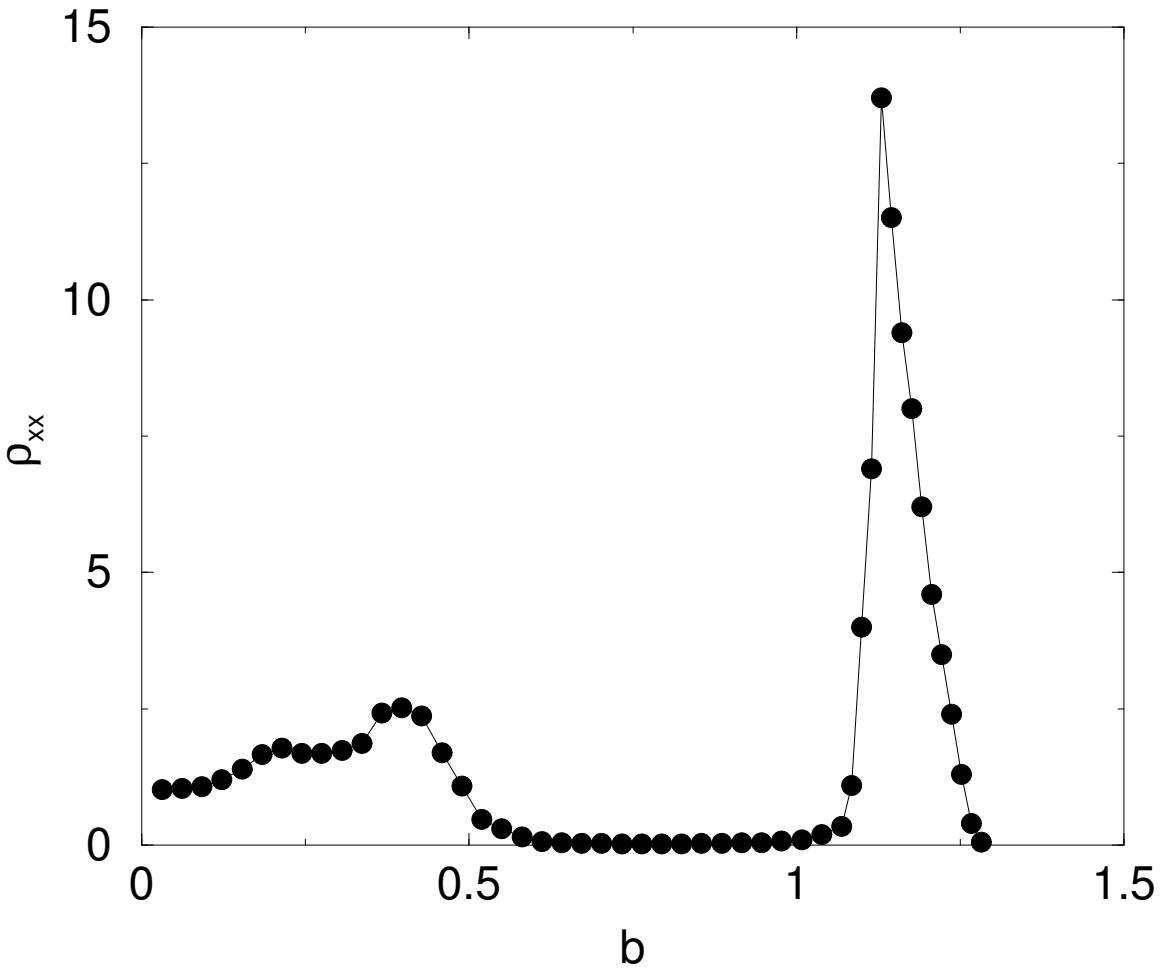

FIG. 1: Numerical results for the longitudinal resistivity, $\rho_{xx}$, as a function of the reduced magnetic field $b$ in units of the zero field resistivity $\rho_0$ for $c = 0.05, \alpha = 0.025$.

Generally, depending on the values of the two small parameters introduced above, $c$ and $\alpha$, the longitudinal resistance as a function of magnetic field may have one, two, or three maximums. A typical curve with three maximums is presented in Fig. 1. We believe to have a qualitative understanding of these features [12] as the result of interplay between two mechanisms similar to those proposed in Refs.[7, 8] leading to successive chaotic or localized electron motion. Here, we focus on the most surprising observation: the appearance of a sharp narrow peak just below the percolation threshold. We introduce the dimensionless magnetic field, $b = B/B_c$, where the characteristic magnetic field $B_c$ corresponds to the condition $NR^2 = 0.36$, so that $R = 0.6N^{-1/2}b^{-1}$. The actual percolation threshold corresponds to $N(R+d/2)^2 = 0.36$, i.e. to a value of $b = b_p$ slightly greater than 1:

$$b_p = (1 - 0.94\sqrt{c})^{-1} \quad (3)$$

It should be noted that near the percolation threshold $\beta = \omega_c\tau \sim \ell/R \sim c^{-1/2}\alpha^{-2} >> 1$. This justifies our method of considering the diffusion of the center of the cyclotron orbit.

To understand the simulation results, we first introduce two more characteristic magnetic fields, $b_1 = \sqrt{c}$ and $b_2 = \alpha/\sqrt{c}$. The first one corresponds to the condition $NRd \sim 1$, so that at $b >> b_1$ the probability of a collision during one cycle, $1 - \exp(-2\pi NRd) \sim \sqrt{c}/b$, is small and most of the electrons perform an unperturbed cyclotron motion. The second one correponds to the condition $\delta \sim d$, where $\delta \sim \alpha R$ is the typical shift of the cyclotron orbit due to scattering during one period. For $b >> b_2$ we have $\delta << d$, which means that motion of the cyclotron orbit center is quasi continuous. The first characteristic field is always below the percolation threshold ($b_1 << 1$) provided that the system of scatterers is dilute, $c << 1$. We will assume that the same is true for the second characteristic field, $b_2 << 1$, which requires the following condition to be fulfilled:

$$\alpha << \sqrt{c}. \qquad (4)$$

The relation between $b_1$ and $b_2$ may be arbitrary [13].

We will consider the interval of magnetic fields such that $b_1, b_2 << b \lesssim 1$. Since in this region $\delta << d$, the motion of the cyclotron orbit center is almost adiabatic, an effective potential energy for this motion can be introduced by averaging the real potential energy of the scattering system over one cyclotron orbit with a given center, as it was shown for the case of weak random potential in Ref. [7]. In the adiabatic approximation the orbit center follows the equipotential lines of the effective potential energy. The specific feature of a diluted system is that, when $b >> b_1$, it is very improbable for an electron to encounter three or more scatterers on the same cyclotron circle.

Fig. 2a shows the simulated electron trajectories in the vicinity of two scatterers. Fig. 2b presents the equipotential lines for the effective potential produced by two scatterers and different types of the orbit center trajectories. The effective potential is non zero within intersecting rings formed by circles of radiuses $R - d/2$ and $R + d/2$ drawn around each scattering center. Examples of trajectories for $\alpha = 0.025$, $c = 0.2$ (satisfying Eq.(4)) are given in Fig. 2c. Within the adiabatic approximation the distance between the orbit center and the scattering center remains constant for all successive collisions. If this distance lies between $R$ and $R + d/2$ the orbit center drifts around the scatterer in the same direction as the electron itself, and the trajectories are restricted to the exterior of any finite cluster. For distances between $R - d/2$ and $R$ the drift goes in the opposite direction, these "interior" trajectories go around holes between the two rings presented in Fig. 2c, and are generally of small size, on the order of $R$.

By definition, at the percolation threshold ($b = b_p$) there exists a single infinite path consisting of arcs of radius $R + d/2$ surrounding the scattering centers, while all the other finite size trajectories correspond to localized motion. At lower fields, $R$ increases and a similar infinite path consists of arcs of radius between $R + d/2$ and $R$. For $b = 1$ the radius of arcs forming an infinite equipotential line becomes equal to $R$. At still lower fields infinite paths cease to exist. Thus infinite equipotential lines of the effective potential exist only in the narrow magnetic field interval $1 < b < b_p$. Outside of this interval all equipotential lines are closed and finite. This observa-tion is key to understanding the origin of the giant peak in Fig. 1.

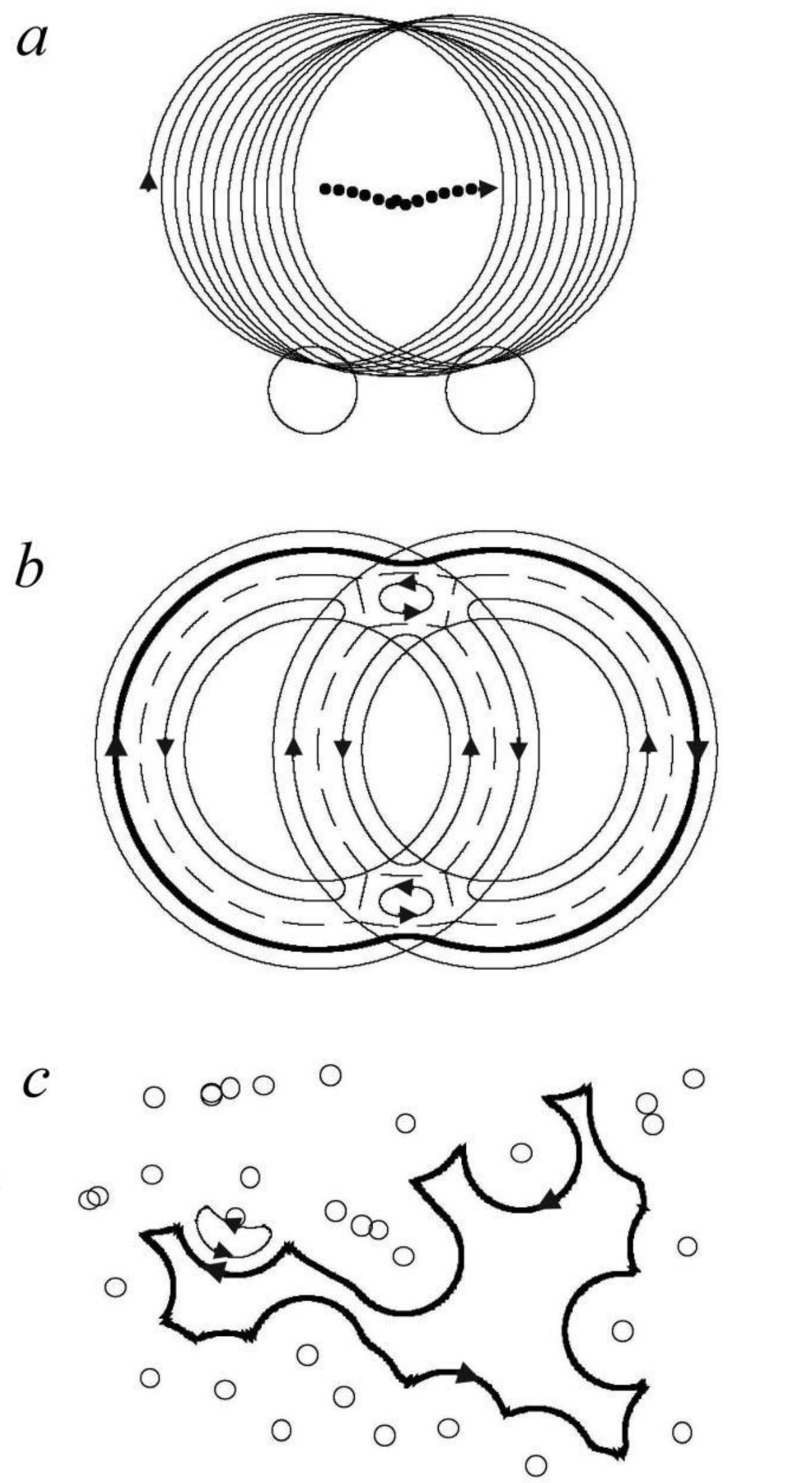

FIG. 2: $a$ - part of a simulated electron trajectory passing from the left scatterer to the right one, the dots show the successive positions of the orbit center. $b$ - Equipotential lines of the effective potential for the two scatterers in $a$. Arrows show the direction of the orbit center drift, dashed lines represent the separatrices. Note the difference between exterior (heavy line) and interior trajectories, as well as the traps at the intersections of the two rings. $c$ - Examples of a simulated exterior and interior trajectories of the orbit center for $\alpha = 0.025, c = 0.2$. Only exterior trajectories may form an infinite path.

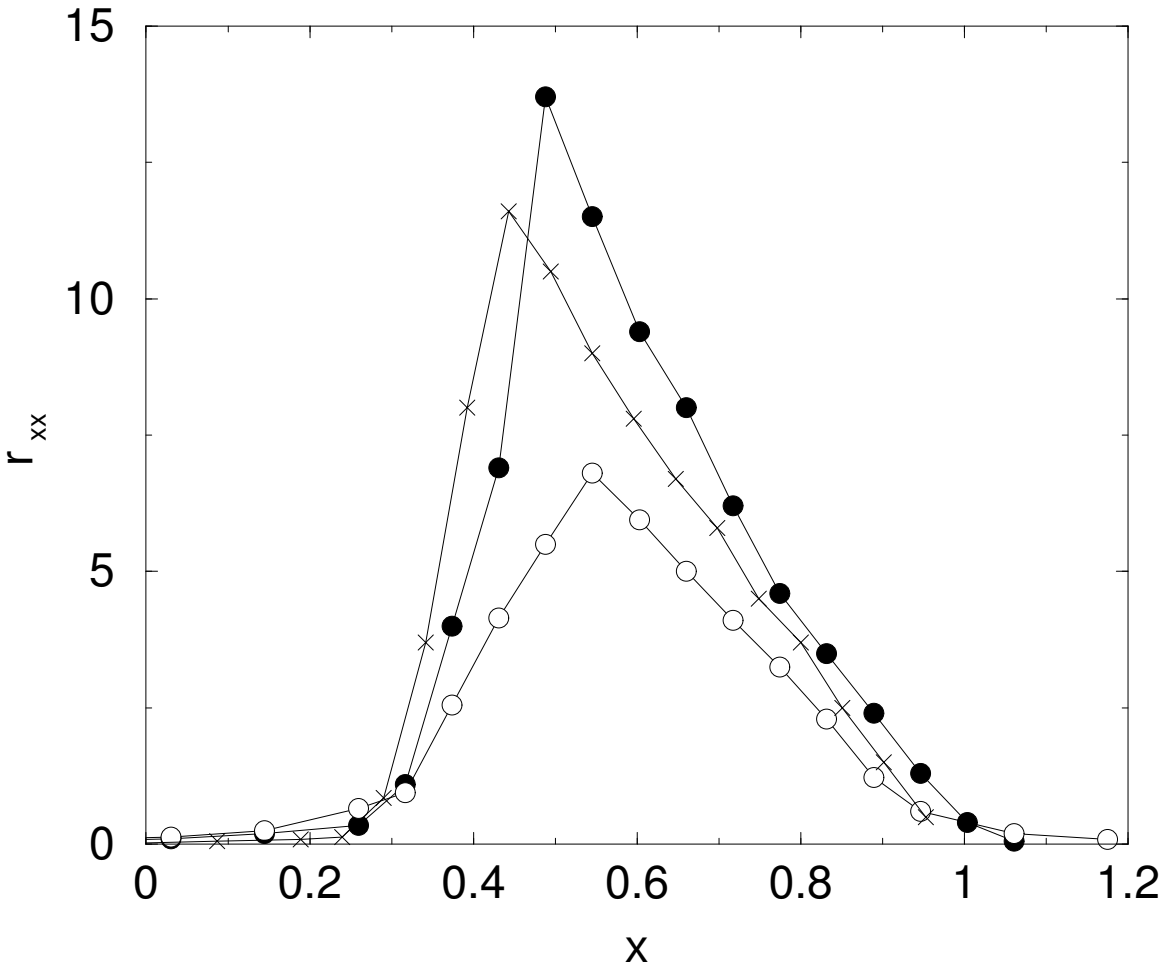

FIG. 3: Longitudinal resistivity curves in the vicinity of the percolation threshold versus $\xi = (b-1)/(b_p-1)$ for several values of the parameters $c$ and $\alpha$. Filled circles correspond to $c = 0.05$, $\alpha = 0.025$, open circles correspond to $c = 0.05$, $\alpha = 0.05$ and crosses correspond to $c = 0.1$, $\alpha = 0.025$.

Since the ratio $\delta/d$ is finite, the motion is not strictly adiabatic (see Fig. 2a). Therefore infinite paths existing in the magnetic field interval $1 < b < b_p$ have a certain "width" (large closed loops, like the one presented in Fig. 2c, are connected by bridges due to non-adiabatic effects) and the averaged diffusion coefficient for the orbit center is also finite.

The magnetic field range where the delocalized motion exists, $b_p - 1$, is proportional to $\sqrt{c}$, see Eq.(3). To check our main idea, that the peak is related to the existence of infinite equipotential lines in the interval $1 < b < b_p$, we plot in Fig. 3 the numerical results for the magnetoresistance for several values of $c$ and $\alpha$ as a function of the parameter $\xi = (b-1)/(b_p-1)$. One can see that indeed the peak always nicely fits into the interval $0 < \xi < 1$, although the peak amplitude varies, decreasing as the characteristic fields $b_1$ and $b_2$ increase and the condition $b_1, b_2 << b_p$ becomes violated.

If we consider the adiabatic motion along the infinite equipotential line as a diffusive motion with a step $\sim R \sim 1/\sqrt{N}$ and a correlation time $\sim R/(\alpha v)$, where $\alpha v$ is the typical drift velocity of the orbit center around a given scatterer, we get a very large, compared to its Boltzmann value, diffusion coefficient $D_{xx}/D_{xx}^B \sim \alpha^{-1}c^{-1/2}$. However, for $b_1 << 1$ only a fraction $\sim \sqrt{c}$ of all electrons can experience collisions, while the majority of them are circling in free space and do not contribute to diffusion. Taking this into account and supposing for a moment that *all* of the "wandering" electrons participate in the diffusive motion, we obtain the peak value $\rho_{xx} \sim 1/\alpha >> 1$. In reality, only a small part of wandering electrons, which happen to be within the "width" of the infinite equipotential line are not localized, so the peak value should be reduced by some factor depending on the degree of non-adiabaticity, $\delta/d$. At present, we are unable to estimate this factor and thus we cannot explain the peak amplitude, nor the small (but finite) value of the resistance to the left of the peak.

In summary, we have studied numerically classical magnetoresistance for electrons in a soft billiard (dilute system of weak short-range scatterers in two dimensions), and we have observed a giant peak in the vicinity of the percolation threshold, at which the longitudinal resistance vanishes. By considering the potential landscape for the effective potential governing the quasi- adiabatic motion of the cyclotron orbit center, we have reveiled a peculiar situation, in which infinite equipotential lines exist only within a narrow magnetic field interval below the percolation threshold. We have shown that the peak corresponds to this interval. The magnitude and the shape of the peak still remain to be understood by properly taking account of non-adiabatic effects.

We thank Boris Shklovskii for valuable comments.